\documentclass[epj]{svjour}
\DeclareUnicodeCharacter{2032}{\ensuremath{^{\prime}}}
\usepackage{graphics}

\usepackage[dvipsnames]{xcolor}
\usepackage{graphicx,color}
\usepackage{enumitem}
\usepackage{cite}
\usepackage{braket}
\usepackage{slashed}
\usepackage[utf8]{inputenc}
\usepackage[
    colorlinks=true,
    linkcolor=blue,
    citecolor=blue,
    urlcolor=blue
]{hyperref}
\usepackage{textcomp}
\usepackage{float}
\usepackage{caption}
\usepackage{subcaption}
\usepackage{makecell}
\usepackage{epsfig}
\usepackage{amsmath,amssymb}
\usepackage[normalem]{ulem}
\usepackage{tikz}

\usetikzlibrary{shapes.geometric, arrows}
\hypersetup{
	colorlinks=true,
	citecolor=black,
	filecolor=black,
	linkcolor=blue,
	urlcolor=blue
}

\begin{document}

\title{A Circle That Won't Return: The Fate of RR Fluxes and D-branes in Type 0A Tachyon Condensation}

\author{Ahmed Rakin Kamal\inst{1,2}
}                     

\institute{
Department of Theoretical Physics and Astrophysics, Masaryk University, Kotlářská 2, CS-61137 Brno, Czechia.
\and 
Department of Mathematics and Natural Sciences, BRAC University, Kha 224, Bir Uttam Rafiqul Islam Avenue, Dhaka 1212, Bangladesh.
}

\mail{ahmedrakinkamaltunok@gmail.com}

\abstract{We study the closed-string tachyon and doubled Ramond--Ramond sector of type 0A in light of the proposed M-theory description on $S^1\vee S^1$-the wedge of two circles joined at a point. In this picture the two RR copies are associated with the two circle components, which we call branches, and tachyon condensation corresponds to shrinking one branch to the type IIA endpoint. From the type 0A equations of motion, we derive the branch-balance condition for a tachyon stationary point and identify the branch-odd RR fluctuation that sources the tachyon around a symmetric background. We then analyze the fate of the collapsing branch RR data as one branch of the wedge collapses to the type IIA endpoint. For an isolated unscreened collapsing-branch $D_p^-$ source, a Gauss-law estimate shows that the long-range RR field-energy cost scales inversely with the shrinking circle and thus becoming infinitely costly, generalizing the $D0^-$-brane decoupling in the original wedge picture. We describe the infrared screening of the relative RR field through an effective higher-form Stückelberg mechanism and distinguish this from a possible discharge of localized relative charge. Finally, using standard Wess--Zumino couplings of D-branes, we identify an effective relative-charge carrier and derive a parametric thin-wall criterion for when such a discharge channel $D_p^- \to D_p^+$ can be energetically favored.
}

\titlerunning{A Circle That Won't Return: RR Fluxes and D-branes in Type 0A Tachyon Condensation}
\maketitle

\section{Introduction}
Dualities have been one of the central reasons for string theory to emerge as a leading candidate for quantum gravity. The fact that apparently different perturbative theories can describe the same physics-for example through T-duality, where compactification on a circle of radius $R$ is equivalent to compactification on a circle of radius $\alpha'/R$, or through S-duality, where a strongly coupled description is mapped to a weakly coupled one has been one of the highlights of the superstring revolution. Historically, T-duality \cite{Buscher:1987qj,Buscher:1987sk,Giveon:1994fu} alongside strong-weak coupling duality and its unification with T-duality into U-duality led to the modern picture in which the five ten-dimensional superstring theories are apparently different limits of a single underlying theory \cite{Font:1990nt,Sen:1994fa,Hull:1994ys,Witten:1995ex}. Dualities have since then become both a conceptual and calculational engine of string theory: they relate strongly coupled regimes to weakly coupled descriptions, clarify the role of D-branes as Ramond-Ramond charged objects, and underlie developments such as mirror symmetry, black-hole microstate counting, holographic gauge/gravity duality, and constraints the higher-derivative corrections to classical supergravity action \cite{Polchinski:1995mt,Candelas:1990rm,Strominger:1996sh,Maldacena:1997re,Hohm:2014eba,Hsia:2024kpi,Wulff:2021fhr,Aggarwal:2025lxf,Ameri:2025bei,Wulff:2024ips}. Furthermore, dualities also seem to help in the story-telling of the early universe \cite{Brandenberger:1988aj,Veneziano:1991ek,Meissner:1991zj,Agrawal:2020xek,Sen:1991zi}. 

The situation has been far less complete for the non-supersymmetric strings. For example, Type 0A/0B and non-supersymmetric heterotic strings have been known for a long time \cite{Dixon:1986iz,Seiberg:1986by,Alvarez-Gaume:1986ghj}. Perturbatively, type $0A$ on a circle of radius $R$ is T-dual to type 0B on a circle of radius $
\alpha'/R$ \cite{Bergman:1999km,Meessen:2001cq}. There were also early attempts to understand type 0 strings from an eleven-dimensional point of view, for example through M-theory on antiperiodic circle \cite{Bergman:1999km}. However, these ingredients did not give a duality web as rigid as the supersymmetric one. The recent proposal of Baykara, Dudas and Vafa (BDV) \cite{Baykara:2026bdv} gives a sharp new picture: type 0A is described by M-theory on the quantum wedge $S^{1}\vee S^{1}$, where different fields probe different effective resolutions of the singular space; the doubled RR sector is branch data, and the closed-string tachyon is the branch-odd radius mode. This has led to a rapidly growing non-supersymmetric duality program relating type 0 orientifolds, heterotic theories, bosonic strings, wedge singularities and cobordism junctions \cite{Baykara:2026web,Altavista:2026brr,Altavista:2026edv,Altavista:2026evd,Dasgupta:2026maq,Basile:2026trt}.

Even with these dualities, the status of non-supersymmetric strings faces a basic difficulty: the simplest examples contain a closed-string tachyon. It is useful to contrast this with the open-string tachyon, whose physical meaning is much better understood. In Sen's picture of open-string tachyon condensation, the tachyon on an unstable D-brane or on a brane--antibrane pair parametrizes the decay of the original brane system leaving just the closed-string vacuum \cite{Sen:1998tt,Sen:1998sm,Witten:1998cd,Sen:2002nu,Sen:2004nf,Sen:1999xm}. This led to the study of brane-antibrane potentials, and also motivated the brane-antibrane inflationary scenario, where the inter-brane separation plays the role of the inflaton and the final tachyonic instability ends inflation \cite{Burgess:2001fx,Cicoli:2024bwq,Chakraborty:2026hob}. On the other hand, closed-string tachyons are more dangerous \cite{Adams:2001sv,Adams:2005rb}, since they signal an instability of the spacetime background itself. The BDV wedge proposal gives a concrete interpretation for the type $0A$ closed-string tachyon: it is the relative size of the two M-theory branches, and its condensation removes one RR branch by sending the corresponding KK/D0 tower to infinite mass. 

In this work we analyze the BDV picture from the type 0A point of view. Type 0A contains a doubled RR sector, and in the $S^1\vee S^1 $ interpretation these two copies are naturally associated with the two branches of the wedge. We first ask how these RR fields enter the tachyon equation of motion. This gives a local condition for when the tachyon can remain at the symmetric point, and shows explicitly how an imbalance between the two RR sectors can source the tachyon. We then turn to the proposed tachyon-condensation endpoint, where one branch of the wedge shrinks and the theory becomes ordinary type IIA on the surviving circle. Since type IIA has only one RR sector, any RR flux or D-brane charge associated with the disappearing branch must have a definite low-energy fate. We therefore study the energy cost of an isolated wrong-branch $D_p^-$ source, describe how the corresponding relative RR field may be screened in the infrared, and distinguish this from the stronger possibility of discharge, in which the localized relative charge is transferred through an additional charge-carrying channel. The goal is to isolate, within the low-energy type 0A description, the constraints that any $0A\to IIA$ endpoint must satisfy for the doubled RR sector and its charged branes. To summarise, our results are
\begin{enumerate}[label=(\alph*)]
    \item the tachyon equation gives a local branch-balance condition and around a \(Q\)-symmetric background, the \(Q\)-odd RR perturbation sources the tachyon but not the metric/dilaton at first order. 
    \item The electric wrong-branch charge has energy \(\mathcal E_-\sim 1/R_-\) and thus gives a way to decouple wrong branch branes. 
    \item  An infinitely massive relative RR field screens long-range relative charge and hence only one of the RR sector appears in type IIA. 
    \item A WZ/DBI fluxed carrier has the correct relative charge for \(D_p^-\to D_p^+\), and if a finite-tension wall exists, the thin-wall barrier is parametrically lowered as the branch collapses and thus makes such a transition energetically favourable. 
\end{enumerate}

\section{Type 0A}
Type $0A$ string theory may be defined directly in the RNS formulation by replacing the supersymmetric type-II GSO projection with a non-supersymmetric one \cite{Dixon:1986iz,Seiberg:1986by}. In the NS sector, \(NS_+\) denotes the GSO-even sector containing the massless excitations, whereas \(NS_-\) contains the NS ground state and therefore gives a closed-string tachyon. In the Ramond sector, \(R_\pm\) denote the two ten-dimensional chiralities of the Ramond ground state. The type \(0A\) Hilbert space is
\begin{equation}
\mathcal H_{0A}
=
(NS_+,NS_+)\oplus (NS_-,NS_-)
\oplus (R_+,R_-)\oplus (R_-,R_+).
\end{equation}
Thus \((NS_+,NS_+)\) gives the universal NS-NS fields i.e $g_{\mu\nu},B_{\mu\nu},\Phi$, while \((NS_-,NS_-)\) gives the closed-string tachyon \(T\). The mixed NSR and RNS sectors are absent, so there are no spacetime fermions. The two RR sectors \((R_+,R_-)\) and \((R_-,R_+)\) give two independent copies of the type-IIA RR potentials, which we denote by \((A_1,C_3)\) and \((A'_1,C'_3)\). Equivalently, type \(0A\) can be obtained as the orbifold of type IIA
\begin{equation}
0A=IIA/(-1)^{F_s},
\end{equation}
where \((-1)^{F_s}\) is spacetime fermion number. In this description, the ordinary type-IIA bosonic fields form the untwisted sector, while the tachyon and the second RR copy arise from the twisted sector.

The orbifold construction implies a \(\mathbb Z_2\) quantum symmetry \(Q\) \cite{Vafa:1989ih}, acting with eigenvalue \(+1\) on untwisted states and \(-1\) on twisted states. Therefore
\begin{align}
Q:\qquad
&T\to -T,\qquad
A_1\to A_1,\qquad C_3\to C_3,\qquad \\
&A'_1\to -A'_1,\qquad C'_3\to -C'_3 .
\end{align}
Introducing the linear combinations
\begin{equation}
A_1^\pm=\frac12(A_1\pm A'_1),\qquad
C_3^\pm=\frac12(C_3\pm C'_3),
\end{equation}
the action of \(Q\) becomes
\begin{equation}
Q:\qquad
T\to -T,\qquad
A_1^+\leftrightarrow A_1^-,
\qquad
C_3^+\leftrightarrow C_3^- .
\end{equation}
Type \(0A\) also admits the worldsheet parity symmetry \(\Omega\), which exchanges the left and right movers. Since the spectrum contains both \((R_+,R_-)\) and \((R_-,R_+)\), this is a symmetry of the full type \(0A\) theory. On the NSNS sector,
\begin{equation}
\Omega:\quad
g_{\mu\nu}\to g_{\mu\nu},\quad
\Phi\to\Phi,\quad
T\to T,\quad
B_2\to -B_2 ,
\end{equation}
whereas in the RR sector \(\Omega\) exchanges the two copies, \(A_1\leftrightarrow A'_1\) and \(C_3\leftrightarrow C'_3\). Equivalently, in the \(A^\pm,C^\pm\) basis it acts diagonally,
\begin{equation}
\Omega:\qquad
A_1^\pm\to \pm A_1^\pm,\qquad
C_3^\pm\to \pm C_3^\pm .
\end{equation}
Thus \(Q\) is the orbifold quantum symmetry, odd on the tachyon and exchanging the two RR branches in the \(\pm\) basis, while \(\Omega\) is worldsheet orientation reversal, even on \(T\), odd on \(B_2\), and diagonal on \(A^\pm_1,C^\pm_3\). Now, the minimal two-derivative string-frame action relevant for our discussion is \cite{Dixon:1986iz,Klebanov:1998yya}
\begin{align}
S&=\frac{1}{2\kappa_{10}^2}\int d^{10}x\sqrt{-g}\Bigg[\int e^{-2\Phi}\left(R+4(\partial\Phi)^2-\frac12(\partial T)^2-V(T)\right)
\nonumber \\
&-\frac12\sum_{p=2,4}\sum_{\sigma=\pm} f_{p,\sigma}(T)|F_p^\sigma|^2
\Bigg].
\label{eq:string-action}
\end{align}
The functions $f_{p,\sigma}(T)$ encode the local tachyon coupling to the doubled RR sector.  The Type 0 quantum symmetry Q gives \footnote{We have kept $\alpha_p$ and $\beta_p$ undermined as these are constants depending on convention. However, the requirements of T-duality can actually fix these coefficients \cite{Klebanov:1998yya,Meessen:2001cq}. }
\begin{equation}
 f_{p,+}(T)=f_{p,-}(-T).
\end{equation}
Therefore, near $T=0$,
\begin{align}
 f_{p,+}(T)&=1+\alpha_pT+\beta_pT^2+\mathcal{O}(T^3),
 \nonumber\\
 f_{p,-}(T)&=1-\alpha_pT+\beta_pT^2+\mathcal{O}(T^3).
 \label{eq:f-expansion}
\end{align}
The potential is even under $T\to -T$,
\begin{equation}
 V(T)=V(0)+\frac12V''(0)T^2+\mathcal{O}(T^4),
 \qquad V''(0)<0
 \label{eq:potential-expansion}
\end{equation}
Varying the action, one arrives at the equation of motion for the tachyon 
\begin{equation}
\nabla_M\left(e^{-2\Phi}\partial^M T\right)-e^{-2\Phi}V'(T)-\frac12\sum_{p=2,4}\sum_{\sigma=\pm}f'_{p,\sigma}(T)|F_p^\sigma|^2=0.
\label{eq:eom-tachyon}
\end{equation}
The last term will be central in the subsequent discussions as evidently the RR fluxes can act as sources for tachyon and can shift the condition for a tachyon stationary point. We analyze this flux-induced source term in section \ref{section 4}. 
\section{ M-theory on $S^1 \vee S^1$} \label{section 3}
The proposed M-theory origin of type \(0A\) is not an ordinary compactification on a
smooth circle, but rather on the singular one-dimensional space
\(S^1_+\vee S^1_-\), the wedge sum of two circles meeting at a point
\cite{Baykara:2026bdv}. The motivation is simple and beautiful but quite rigid. Type \(0A\)
contains two Ramond--Ramond one-forms and two Ramond--Ramond three-forms, and
hence suggests two independent M-theory circle-like sectors, while a genuine product
\(S^1\times S^1\) would give the wrong spacetime dimension. The wedge geometry
provides a minimal way of having two circle branches without introducing an extra
macroscopic dimension. The exchange of the two circles is then identified with the
type \(0A\) quantum symmetry \(Q\), under which the tachyon is odd and the two RR
branches are exchanged. In this language the tachyon is naturally interpreted as a
branch-odd combination measuring the relative size of the two circles, while the
branch-even combination is the dilaton.

A crucial point is that \(S^1_+\vee S^1_-\) should not be understood as an ordinary
classical space on which all eleven-dimensional fields obey the same boundary
condition. Rather, different fields are assigned different effective resolutions of
the singular wedge. In the disconnected resolution property, or DRP, the node is
resolved as \(S^1_+\cup S^1_-\), so that a field may live independently on the two
branches; this is the appropriate behavior for fields which are doubled in the
ten-dimensional type \(0A\) spectrum. In the connected resolution property, or CRP,
the two branches are instead resolved into a single connected circle, with the two
branch values identified at the contact point. One must also allow the
orientation-reversed connected resolution, CRP\('\), since type \(0A\) may be viewed
as arising from either IIA or IIA\('\). Fields which are smooth with respect to both
the disconnected and connected resolutions are said to have the strong smoothness
property, SSP. Thus DRP fields carry branch-localized Kaluza--Klein data, while SSP
fields feel only the connected, branch-symmetric geometry.

With these assignments, the type \(0A\) field content is reproduced in a rather
geometric way from M-theory. The ten-dimensional metric \(g_{\mu\nu}\) is an SSP field, since
there is only one spacetime metric. By contrast, the metric components
\(g_{\mu +}\) and \(g_{\mu -}\) are DRP fields and become the two RR one-forms
\(A^+_\mu\) and \(A^-_\mu\). Similarly, the branch radii \(R_+^2\) and \(R_-^2\)
are DRP data whose even and odd combinations are interpreted as the dilaton and
tachyon. For the eleven-dimensional three-form the assignment is reversed:
components with one leg along the wedge give a single NSNS two-form \(B_{\mu\nu}\)
and are SSP, whereas the components entirely along the ten-dimensional spacetime
are DRP and give the doubled RR three-forms \(C^+_{\mu\nu\rho}\) and
\(C^-_{\mu\nu\rho}\). Finally, the gravitino is assigned an odd SSP-type condition,
usually denoted SSP\(^*\), so that the would-be constant fermionic zero mode is
removed, in agreement with the absence of perturbative spacetime fermions in type
\(0A\). The resulting picture is therefore intrinsically quantum-geometric i.e the
wedge is not resolved once and for all, but is probed differently by different
components of the eleven-dimensional fields as seen in the table \ref{tab:wedge-summary}. 
\begin{table}[t]
\centering
\footnotesize
\setlength{\tabcolsep}{3pt}
\renewcommand{\arraystretch}{1.05}
\begin{tabular}{@{}>{\raggedright\arraybackslash}p{0.31\linewidth}>{\raggedright\arraybackslash}p{0.60\linewidth}@{}}
\hline
11d data & Type \(0A\) field and resolution \\
\hline
\(G_{\mu\nu}\) & \(g_{\mu\nu}\); SSP, single metric. \\
\(G_{\mu\pm}\) & \(A^\pm_\mu\); DRP, doubled RR 1-forms. \\
\(G_{\pm\pm}\sim R_\pm^2\) & even/odd radius data: \(\Phi,T\); DRP. \\
\(C^{(11)}_{\mu\nu\pm}\) & \(B_{\mu\nu}\); SSP, single NSNS 2-form. \\
\(C^{(11)}_{\mu\nu\rho}\) & \(C^\pm_{\mu\nu\rho}\); DRP, doubled RR 3-forms. \\
\(\Psi_M\) & no massless spacetime fermion; SSP\(^*\). \\
\hline
\end{tabular}
\caption{Summary of the \(S^1_+\vee S^1_-\) field assignments of \cite{Baykara:2026bdv}. DRP denotes disconnected resolution property, and SSP denotes strong smoothness property.}
\label{tab:wedge-summary}
\end{table}

The D0-brane sector gives one of the sharpest pieces of evidence for the
\(S^1_+\vee S^1_-\) interpretation of type \(0A\). In ordinary type IIA, the
D0-brane is the Kaluza--Klein excitation of eleven-dimensional momentum along the
M-theory circle. The proposed \(0A\) scenario replaces this single circle by two
branches, and hence predicts two elementary KK charges. These are naturally
identified with the two stable \(0A\) D-particles \(D0^+\) and \(D0^-\), charged
under the two RR one-forms \(A^+_\mu\) and \(A^-_\mu\). Thus \(D0^+\) is the
lightest unit of momentum on \(S^1_+\), while \(D0^-\) is the lightest unit of
momentum on \(S^1_-\).

The tachyon then has a particularly transparent geometric meaning. It is the branch-odd radius modulus, measuring the relative size of the two components of the wedge. In the normalization suggested by the \(D0^\pm\) tensions, the masses take the form \cite{Garousi:1999fu}
\begin{equation}
m_{D0^\pm}
=
\frac{M_s}{\sqrt{2}\lambda_s}
\frac{1}{\sqrt{1\pm T/2}} .
\end{equation}
Since a KK mass scales as \(m_{\rm KK}\sim 1/R\), this immediately suggests
\begin{equation}
R_\pm^2=
\frac{2\lambda_s^2}{M_s^2}
\left(1\pm \frac{T}{2}\right),
\qquad
\frac{T}{2}
=
\frac{R_+^2-R_-^2}{R_+^2+R_-^2}.
\end{equation}
Thus the point \(T=0\) is the symmetric wedge \(R_+=R_-\), while the endpoints
\(T=\pm2\) are not small perturbations of type \(0A\), but finite-distance limits in which one of the two circles collapses \cite{Baykara:2026bdv}. In particular, for
\(T=+2\) one has \(R_-=0\), and hence the corresponding KK mass becomes infinitely massive,
\begin{equation}
m_{D0^-}\to\infty ,
\end{equation}
whereas \(D0^+\) remains at finite mass and is naturally identified with the
ordinary type IIA D0-brane. The endpoint \(T=-2\) gives the same physics with the
two branches exchanged.

In this sense, type \(0A\) tachyon condensation is geometrized as the collapse of
one branch of the figure-eight. The unstable \(0A\) point corresponds to the
quantum wedge with two equal circles, while the endpoint is ordinary M-theory on
the surviving circle, namely type IIA. This picture also explains why the
disappearing branch does not lead to an infinite tower of light states as one would assume from the distance conjecture  \cite{Ooguri:2006in}: its KK
modes become heavy rather than light as \(R_\pm \to0\). Therefore the important point to notice is that \(0A\to IIA\)
transition should not be understood as decompactification, but as a finite-distance
removal of one RR branch. The doubled RR sector of type \(0A\) should thereby reduced
to the single RR sector of type IIA, while the surviving D0-brane becomes the
usual KK mode of the remaining M-theory circle.

\section{Perturbation around a \(Q\)-symmetric background}\label{section 4}

We now use the tachyon equation of motion eq. \eqref{eq:eom-tachyon} to study how the doubled RR sector affects the tachyon dynamics. The important term is the last one in eq. \eqref{eq:eom-tachyon}, since the RR kinetic functions \(f_{p,\sigma}(T)\) depends on the tachyon. Thus RR fluxes do not merely contribute to the stress tensor; they can also act as direct sources for \(T\), and can shift both the location and the effective mass of a tachyon stationary point.

Let us first recall what we mean by a \(Q\)-symmetric background. The type \(0A\) quantum symmetry \(Q\) acts by \(T\to -T\) and exchanges the two RR branches, \(F_p^+\leftrightarrow F_p^-\). A \(Q\)-symmetric configuration is therefore one in which the tachyon sits at the fixed point \(T=0\), and the two RR sectors are chosen symmetrically. The choice for this is thus
\begin{equation}
  T=0,\qquad F_p^+=F_p^- .  
\end{equation}
More generally, a background can remain stationary at \(T=0\) only if the RR contribution to the tachyon equation cancels between the two branches. This is the branch-balance condition derived below. Somewhat similar observation was made in\cite{Klebanov:1998yya}, in the context of type \(0B\), that RR fluxes can shift the tachyon mass. Here we will use the same basic mechanism in type \(0A\), but organize it in the branch language suggested by the \(S^1_+\vee S^1_-\) picture. This will allow us to distinguish the branch-even RR background from branch-odd perturbations which measure an imbalance between the two RR sectors.
 Consider a fixed background with $T=0$.  Since $V'(0)=0$ and
\begin{equation}
 f'_{p,+}(0)=\alpha_p,
 \qquad
 f'_{p,-}(0)=-\alpha_p,
\end{equation}
we find from eq. \eqref{eq:eom-tachyon}
\begin{equation}
 -\frac12 e^{2\Phi}\sum_{p=2,4}\alpha_p\left(|F_p^+|^2-|F_p^-|^2\right)=0.
\end{equation}
Thus a fixed background with $T=0$ is stationary if 
\begin{equation}
 \sum_{p=2,4}\alpha_p\left(|F_p^+|^2-|F_p^-|^2\right)=0.
 \label{eq:branchbalance}
\end{equation}
This is the local branch-balance condition in $S^1 \vee S^1$ picture as the two fluxes live on the two circles. Around a genuinely $Q$-symmetric RR background,
\begin{equation}
 T=0,
 \qquad
 F_p^+=F_p^-=K_p,
 \label{eq:symmetric-background}
\end{equation}
it is automatically satisfied. A small branch-odd perturbation can tell us how things change. We therefore write
\begin{equation}
 F_p^+=K_p+\epsilon G_p,
 \qquad
 F_p^-=K_p-\epsilon G_p,
 \qquad
 T=\epsilon\tau.
 \label{eq:linearization}
\end{equation}
Here $K_p$ is branch-even and $G_p$ is branch-odd. In the picture of \cite{Baykara:2026bdv}, branch-even would correspond to $R_+=R_-$ and branch-odd when they are not equal. So, $G_p$ would correspond to the inequality of the two branches for example in the onset of tachyon condensation. 
\begin{align}
 L_{\rm}^{(2)}
 &=e^{-2\Phi}\left[-\frac12(\partial\tau)^2-\frac12V''(0)\tau^2\right] \nonumber \\
 &-\sum_{p=2,4}\left[|G_p|^2+2\alpha_p\tau K_p\cdot G_p+\beta_p\tau^2|K_p|^2\right].
 \label{eq:L2}
\end{align}
One can immediately see that eq. \eqref{eq:branchbalance} condition becomes
\begin{equation}
   4 \sum_{p=2,4}\alpha_p \epsilon K_p \cdot G_p=0
\end{equation}
If one insists that the perturbed configuration remains on the $T=0$ point, the linearized branchbalance condition requires $\sum_p \alpha_p K_p \cdot G_p=0$. This may be achieved either by cancellation among the $p=2,4$ sectors or by choosing odd RR perturbations orthogonal to the background fluxes. In the full dynamical problem, however, a nonzero $\sum_p \alpha_p K_p \cdot G_p$ simply sources the tachyon fluctuation $\tau$. This is consistent even in the light of \cite{Baykara:2026bdv} where a branch-balance ($R_+=R_-$) is possible when you have the two RR sectors to be exactly equal. One can also notice that the tachyon mass is shifted by
\begin{equation}
 m_{\rm eff}^2=V''(0)+2e^{2\Phi}\sum_{p=2,4}\beta_p|K_p|^2.
 \label{eq:meff}
\end{equation}
The branch-odd perturbation terms (such as $T^2\left|G_p\right|^2$) enter at $\mathcal{O}(\epsilon^4)$. Lastly, for a $p$-form term $-\frac12f(T)|F_p|^2$, the stress tensor is
\begin{equation}
 T_{MN}^{(p)}=f(T)\left[\frac{1}{(p-1)!}F_{M P_2\cdots P_p}F_N{}^{P_2\cdots P_p}-\frac12g_{MN}|F_p|^2\right].
 \label{eq:stress}
\end{equation}
At $T=0$, $f_{p,+}(0)=f_{p,-}(0)=1$.  For
\begin{equation}
 F_p^+=K_p+\epsilon G_p,
 \qquad
 F_p^-=K_p-\epsilon G_p,
\end{equation}
the terms linear in $\epsilon$ cancel between the $+$ and $-$ sectors:
\begin{equation}
 T_{MN}[K_p+\epsilon G_p]+T_{MN}[K_p-\epsilon G_p]=2T_{MN}[K_p]+\mathcal{O}(\epsilon^2).
\end{equation}
The terms linear in $\tau$ from $f_{p,+}$ and $f_{p,-}$ cancel as well.  Therefore the metric and dilaton are not sourced at first order by the $Q$-odd sector.  The tachyon and RR equations define a consistent first-order odd subsystem on a fixed branch-even.

In the interpretation of \cite{Baykara:2026bdv}, the doubled RR fields of type \(0A\) are naturally associated with DRP data on the two components of the wedge \(S^1_+\vee S^1_-\). In the perturbative analysis around a \(Q\)-symmetric point, it is therefore useful to introduce branch-even and branch-odd combinations
\begin{equation}
F_p^{\rm e}=F_p^+ + F_p^-,
\qquad
F_p^{\rm o}=F_p^+ - F_p^- .
\end{equation}
For the linearized ansatz
\begin{equation}
F_p^+=K_p+\epsilon G_p,
\qquad
F_p^-=K_p-\epsilon G_p,
\end{equation}
one has
\begin{equation}
F_p^{\rm e}=2K_p,
\qquad
F_p^{\rm o}=2\epsilon G_p.
\end{equation}
Thus \(G_p\) is the normalized coefficient of the branch-odd RR fluctuation, or equivalently the linearized imbalance between the two DRP RR sectors. This provides a useful diagnostic of how relative RR data couple to the tachyon near a \(Q\)-symmetric background. In particular, the branch-balance condition obtained above says that a configuration can remain at the symmetric point \(T=0\) only when the branch-odd RR source is appropriately balanced. If this condition is not imposed, the branch-odd RR perturbation sources the tachyon fluctuation, while the metric and dilaton remain unperturbed at first order around a \(Q\)-symmetric background.

This observation becomes especially useful when combined with the proposed tachyon-condensation endpoint of \cite{Baykara:2026bdv}. In the normalization
\begin{equation}
\frac{T}{2}
=
\frac{R_+^2-R_-^2}{R_+^2+R_-^2},
\end{equation}
the limit \(R_-\to0\) corresponds to \(T\to2\), and the endpoint is expected to be ordinary type IIA on the surviving circle. We should stress that this does not mean that tachyon condensation requires an RR flux imbalance; the tachyon can condense even when the relative RR sector is absent. Rather, the point is that if relative RR flux or relative RR charge is present during the evolution, the type IIA endpoint has no independent long-range RR field with which to measure it. The fate of such relative RR data is therefore a natural question in the low-energy type \(0A\) description. In the next section we analyze this question from three complementary viewpoints: the Gauss-law cost of an isolated wrong-branch source, the possible screening of the relative RR field, and the additional charge accounting required for a genuine discharge process.

\section{Tachyon Condensation to IIA}\label{section 5}
As discussed in section \ref{section 3}, the proposal of \cite{Baykara:2026bdv} argues that as one of the wedged circles, say for example $R_-$, shrinks to zero, the tachyon condenses at $T=2$ and one transitions to type IIA. At the endpoint, one sees that $m_{D0^-}\rightarrow \infty$. Let's now revisit this picture and see what happens for a generic $D_p-$ branes and $F_p$ RR-forms.
\subsection{Unscreened wrong-branch energy cost}
Consider now the RR-sector action
\begin{equation}
S_{\mathrm{RR}}=-\frac{1}{4 \kappa_{10}^2} \int\left[\sum_{\pm} Z_{\pm}(T) F_{p+2}^{\pm} \wedge * F_{p+2}^{\pm}\right]
\end{equation}
here $Z_{\pm}$ is need not be the same as $f_{\pm}$ as $f_{\pm}$ has been worked out only at $T=0$. Nevertheless, we assume that the kinetic coefficient should scale like the branch $Z_{\pm}\propto R_{\pm}$ \footnote{A way to see the scaling is to start from an eleven-dimensional kinetic term for a branch-supported mode $G^{( \pm)}$,
$$
S_{11, \pm}=-\frac{1}{4 \kappa_{11}^2} \int_{M_{10} \times S_{ \pm}^1} G^{( \pm)} \wedge *_{11} G^{( \pm)}
$$
If the mode is independent of the circle coordinate $y_{ \pm}$, integration over the branch gives
$$
S_{10, \pm}=-\frac{2 \pi R_{ \pm}}{4 \kappa_{11}^2} \int_{M_{10}} G^{( \pm)} \wedge *_{10} G^{( \pm)}
$$
which tells us that $Z_{\pm}\propto R_{\pm}$ upon compactification. One, ofcourse, has to be careful with the compactification statement as the Wedge is supposed to be interpretted as a quantum geometry. However, this process gives a rough sketch to determine the scaling of $Z_-$ in terms of $R_-$.} upto powers that do not effect the main result i.e 
\begin{equation}\label{eq: main scaling}
R_{-} \rightarrow 0 \quad \Rightarrow \quad Z_{-}(T) \rightarrow 0 .
\end{equation}
With this in hand, let us try to derive the energy of an isolated $D_p^{-}$ source which is a direct Gauss-law computation. The $D_p^-$ brane couples to $C_{p+1}$ through
\begin{equation}
 S^{-}=\mu_p^{(0)} \int_{W_{p+1}} C_{p+1}^{-}   
\end{equation}
The equation of motion is
\begin{equation}
d\left(Z_{-} \star F_{p+2}^{-}\right)=2 \kappa_{10}^2 \mu_p^{(0)} \delta_{9-p}\left(W_{p+1}\right) .
\end{equation}
For a static flat brane, let $r$ be the radius in the $9-p$ transverse dimensions. Integrating both sides over a transverse ball $B_n(r)$ gives
\begin{equation}
E_{-}(r)=\frac{2 \kappa_{10}^2 \mu_p^{(0)}}{Z_{-} \Omega_{8-p} r^{8-p}}, \quad \Omega_n=\frac{2 \pi^{(n+1) / 2}}{\Gamma((n+1) / 2)}
\end{equation}
This already shows the essential point i.e $E_{-}(r) \propto \frac{1}{Z_{-}}$. As the branch collapses, a fixed amount of $Q_{-}$charge requires a larger and larger long-range electric field.\footnote{Here \(Q_-\) denotes the electric RR charge of the \(D_p^-\)-brane, measured by the flux
\[
Q_-=\int_{S^{8-p}} Z_-\star F_{p+2}^- .
\]
Thus the Gauss-law estimate is a fixed-electric-charge computation for an isolated unscreened source.} The field energy per unit p-volume is
\begin{equation}\label{Unscreened energy}
\mathcal{E}_{-}=\frac{\kappa_{10}^2\left(\mu_p^{(0)}\right)^2}{Z_{-} \Omega_{8-p}} \begin{cases}\frac{r_c^{p-7}-L^{p-7}}{7-p}, & p<7, \\ \ln \frac{L}{r_c}, & p=7, \\ L-r_c, & p=8 .\end{cases}
\end{equation}
Here $r_c$ is a UV cutoff and $L$ is an IR cutoff. Note that this gives us the important universal factor
\begin{equation}
\mathcal{E}_{-} \propto \frac{1}{R_{-}} .
\end{equation}
The EFT conclusion is that an isolated unscreened wrong-branch RR charge becomes increasingly costly as the $R_-$-branch tends to zero. This should be viewed as the higher- $p$ analogue of the BDV statement that the wrong-branch $D_0^{-}$state becomes infinitely massive when $R_{-} \rightarrow 0$. The immediate implication is not that $D_p^{-}$must decay, but rather that an isolated unscreened $D_p^{-}$cannot remain as a finite-energy low-energy object at the Type IIA endpoint. It may therefore simply decouple from the low-energy spectrum. A stronger possibility, which requires additional dynamics, is that before the endpoint is reached the $D_p^{-}$discharges its relative Type 0A charge through a suitable wall. We analyze what such a discharge channel would have to look like later in the section. 
\subsection{Effective screening of the relative RR field}
Another possible way in which the relative RR sector can disappear from the low-energy endpoint is through a higher-form Stückelberg, or Higgs, mechanism. Before introducing this mechanism, let us first make the charge bookkeeping precise. At the perturbative Type $0 A$ point, the doubled RR sector allows two independent electric charge labels, which we denote by ( $Q_{+}, Q_{-}$). It's useful to define the charge vector (at $T=0$)
\begin{equation}
D_p^{+}: \quad\left(Q_{+}, Q_{-}\right)=(1,0), \quad D_p^{-}: \quad\left(Q_{+}, Q_{-}\right)=(0,1)
\end{equation}
Two linear combinations would be useful for our analysis
\begin{equation}
Q_D=Q_{+}+Q_{-}, \quad Q_R=Q_{-}-Q_{+} .
\end{equation}
Here $Q_D$ is the diagonal charge, and $Q_R$ is the relative charge. In these conventions,
\begin{equation}
\begin{array}{lll}
D_p^{+}: & Q_D=1, & Q_R=-1, \\
D_p^{-}: & Q_D=1, & Q_R=+1 .
\end{array}
\end{equation}
So the two Type 0 branes carry the same diagonal charge and opposite relative charge. An important point is that the Type II endpoint can remember $Q_D$, but it has no room to remember an unscreened $Q_R$ once the doubled RR description has collapsed to a single RR field. 
This observation suggests the following necessary effective description of the endpoint relative sector. Since \(Q_R\) is the charge measured by the relative RR gauge field, thus relative combination is
\begin{equation}
C_R \equiv C_{p+1}^{-}-C_{p+1}^{+}. 
\end{equation}
If $C_R$ remained massless at the $R_{-} \rightarrow 0$ endpoint, an asymptotic observer could still measure the relative flux and hence distinguish $Q_R=+1$ from $Q_R=-1$. This would be incompatible with the interpretation of the endpoint as ordinary Type IIA with a single RR sector. Therefore, in any effective description of the endpoint, $C_R$ must either be projected out, confined, or gapped. A minimal gauge invariant way to parametrize the last possibility is to introduce a $p$-form Stückelberg field $\Lambda_p$ and the gauge-invariant combination
\begin{equation}
\mathcal C_R
=
C_{p+1}^{-}-C_{p+1}^{+}-d\Lambda_p .
\end{equation}
Under the two RR gauge transformations
\begin{equation}
C_{p+1}^{\pm}\to C_{p+1}^{\pm}+d\lambda_p^{\pm},
\end{equation}
we assign
\begin{equation}
\Lambda_p\to \Lambda_p+\lambda_p^{-}-\lambda_p^{+},
\end{equation}
so that \(\mathcal C_R\) is gauge invariant. The corresponding effective relative-sector action may then contain
\begin{equation}
S_R
=
-\frac{1}{4\kappa_{10}^2}
\int
\left[
Z_R\,F_R\wedge \star F_R
+
M_R^2(T)\,\mathcal C_R\wedge \star \mathcal C_R
\right]. 
\end{equation}
Here
\begin{equation}
Z_R=\frac{Z_+Z_-}{Z_++Z_-}
\end{equation}
is the relative kinetic coefficient obtained after diagonalizing the \(C^+\), \(C^-\) kinetic terms. The coefficient \(M_R^2(T)\) is an effective gap parameter for the relative RR mode, encoding the coupling of \(C_R\) to the singular collapsing branch sector. If \(Z_-\to0\) as \(R_-\to0\), then \(Z_R\to0\). Hence the canonically normalized relative mass scales as
\begin{equation}
m_R^2\sim \frac{M_R^2(T)}{Z_R}.
\end{equation}
Consequently, if $M_R^2(T)$ does not vanish faster than $Z_R$, then $m_R^2 \rightarrow \infty$ at the endpoint. This can also be seen via dimensional analysis as the physical gap of a mode associated with the collapsing branch should scale like $m_R \sim \frac{1}{R_{-}}$. In that regime the relative field becomes infinitely short-ranged, and $Q_R$ is no longer measurable as a long-range charge in the Type IIA infrared theory. The massive $C_R$ story is thus an analogue of wrong-branch decoupling while flowing to type IIA. 

It is important to note that this should be understood as an effective parametrization of relative-charge screening, not as a microscopic derivation of a specific junction field in the present construction. This is analogous to the familiar Stückelberg realization of massive gauge fields in string compactifications, where RR axions or higher-form fields can render gauge fields massive through gauge-invariant mass terms \cite{Ibanez:1998qp,Berasaluce-Gonzalez:2013bba,Kuzenko:2020zad}. In the present wedge setting, the microscopic origin of \(\Lambda_p\) is not known; its introduction is motivated only by the fact that branch-point constructions can involve additional degrees of freedom localized at the core of the transition \cite{Altavista:2026evd,Altavista:2026edv,Altavista:2026brr}.

In the diagonal basis, we have
\begin{equation}
    S^{D_p^{-}}=\mu_p^{(0)} \int\left[C_D+\frac{Z_{+}}{Z_{+}+Z_{-}} C_R\right]
\end{equation}
At distances $r \gg m_R^{-1}$ or in other words since $C_R$ is infinitely massive, the relative field is absent, so the object is seen only through the diagonal coupling:
\begin{equation}
S^{D_p^{-}} \rightarrow \mu_p^{(0)} \int C_D .
\end{equation}
and at the endpoint one has type IIA which has a single $D_p$ and a single $C_{p+1}$ gauge field. Basically, the possibilities are such that either the $-$ sector completely decouples (as happens for $D_0^-$ in \cite{Baykara:2026bdv}) or before the endpoint, some screening mechanism takes place when the $R_-$ circle shrinks to zero. In this way, a diagonal version remains in type IIA. Thus, screening alone removes the long-range $C_R$ field, but if the process is to represent a genuine $D_p^{-} \rightarrow D_p^{+}$-transition rather than simple decoupling, one must also identify a discharge channel carrying the relative Type 0A charge.

\subsection{Discharge of $D_p^-$}

It is important to distinguish screening from discharge. A Stückelberg mass for \(C_R\) removes the long-range field that measures \(Q_R\), but it does not by itself erase the localized charge carried by a \(D_p^-\). In a massive gauge theory, charged objects may still exist; their fields are merely short-ranged. Thus the relative Higgsing mechanism is sufficient to explain why \(Q_R\) is absent from the infrared Type IIA spectrum, but it is not sufficient to prove a transition \((0,1)\to(1,0)\). However, if we are interested in the conversion \(D_p^-\to D_p^+\) rather than just screening, something more should be done. A genuine conversion \(D_p^-\to D_p^+\) requires an additional wall or defect carrying
\[
\Delta(Q_+,Q_-)=(+1,-1).
\]
Equivalently, the \(D_p^-\) can either decouple as an infinitely costly wrong-branch state, or it can discharge before the endpoint if such a relative-charge carrier exists.

Let's now look at the discharge mechanism in detail. We start by postulating an effective relative-charge carrier; its microscopic wedge realization remains open. Consider a $D(p+2)$-brane wrapping $\mathbb{R}^{p, 1} \times S^2$ and carrying quantized magnetic flux $\frac{1}{2 \pi} \int_{S^2} F=n \in \mathbb{Z}$. The $D(p+2)$ fluxed shell \footnote{For example, in the type \(0A\) context, such an effective relative carrier could be modeled by a composite \(D(p+2)^+ + \overline{D(p+2)^-}\), with worldvolume flux chosen so that the induced WZ charge is \(+C^+_{p+1}-C^-_{p+1}\), or by a more intrinsically wedge-localized junction object or possibly by cobordism conjecture motivation \cite{Dierigl:2026psj}.} is a candidate realization of the relative wall's charge. The standard D-brane Wess--Zumino coupling provides a natural template such that worldvolume flux on a higher D-brane can induce lower D-brane charge (for example the Myers effect or Brane-flux annihilation \cite{Myers:1999ps,Kachru:2002gs}). The effective Wess-Zumino coupling would be \footnote{In presence of tachyon, the WZ action might be modified, see for example \cite{Kluson:2000iy}. }
\begin{equation}
S_{\mathrm{WZ}}=\mu_{p+2}^{(0)} \int_{\mathbb{R}^{p, 1} \times S^2}\left(C_{p+1}^{+}-C_{p+1}^{-}\right) \wedge\left(2 \pi \alpha^{\prime} F\right)
\end{equation}
Using $\int_{S^2} F=2 \pi n, \quad \mu_p^{(0)}=(2 \pi)^2 \alpha^{\prime} \mu_{p+2}^{(0)}$, this reduces to, 
\begin{equation}
S_{\mathrm{WZ}}=n \mu_p^{(0)} \int_{\mathbb{R}^{p, 1}}\left(C_{p+1}^{+}-C_{p+1}^{-}\right)
\end{equation}
Hence, the shell carries $\Delta Q_{+}=+n, \quad \Delta Q_{-}=-n$ and for $n=1$ 
\begin{equation}
    (0,1)+(1,-1)=(1,0),
\end{equation}
it can provide the relative charge transfer. We can also compute the DBI tension per unit p-volume
\begin{align}
T_{\mathrm{shell}}^{\mathrm{DBI}}(r)&=T_{p+2}^{(0)} \int d \theta d \phi \sqrt{\operatorname{det}\left(g+2 \pi \alpha^{\prime} F\right)}\nonumber \\
&=4 \pi T_{p+2}^{(0)} \sqrt{r^4+\pi^2 \alpha^{\prime 2} n^2}
\end{align}
and in the limit when the $S^2$ shrinks,
\begin{equation}
    T_{\text {shell }}^{\mathrm{DBI}}(0)\simeq 4 \pi^2 \alpha^{\prime}|n| T_{p+2}^{(0)}\simeq|n| T_p^{(0)}.
\end{equation}
Thus the collapsed fluxed shell carries $n$ units of $D_p$-brane tension/charge, as expected from the WZ coupling. This is the standard DBI realization of lower-brane charge dissolved as worldvolume flux \cite{Myers:1999ps}. Thus, our proposed charge-carrying wall energetically consistent with dissolved $D_p$ charge.

This gives us the strength to work with the assumption now that such a finite-tension relative wall exists. The natural Euclidean decay picture on the $D_p^{-}$ worldvolume is the usual thin-wall bounce \cite{Coleman:1977py,Callan:1977pt,Coleman:1980aw}. For $p \geq 1$, the Euclidean worldvolume dimension is $d=p+1,$ so a spherical bubble of the lower-energy phase has boundary $S^p$ of radius $\rho$. Let $T_{\text {wall }}$ be the wall tension and $\Delta \mathcal{T} \equiv \mathcal{T}\left(D_p^{-}\right)-\mathcal{T}\left(D_p^{+}\right)$
the worldvolume energy-density difference. The Euclidean action is
\begin{equation}
B(\rho)=\underbrace{\Omega_p \rho^p T_{\text {wall }}}_{\text {wall cost }}-\underbrace{\frac{\Omega_p}{p+1} \rho^{p+1} \Delta \mathcal{T}}_{\text {volume gain }}, \quad \Omega_p=\frac{2 \pi^{(p+1) / 2}}{\Gamma\left(\frac{p+1}{2}\right)} .
\end{equation}
One finds by extremizing that $\rho_*=\frac{p T_{\mathrm{wall}}}{\Delta \mathcal{T}}$ and substituting back in the on-shell bounce action 
\begin{equation}
B_*=\frac{\Omega_p}{p+1} p^p \frac{T_{\mathrm{wall}}^{p+1}}{(\Delta \mathcal{T})^p}
\end{equation}
The decay rate per unit worldvolume is then
\begin{equation}
\frac{\Gamma}{V_{p+1}} \sim e^{-B_*} .
\end{equation}
From \eqref{Unscreened energy}, we can see that 
\begin{equation}
\Delta \mathcal{T}_{\mathrm{rel}} \sim \mathcal{E}_{-} \sim \frac{1}{Z_{-}} 
\end{equation}
Interestingly,
\begin{equation}
\rho_* \sim Z_{-}, \quad B_* \sim Z_{-}^p .
\end{equation}
Thus for every $p\geq1$ and using \eqref{eq: main scaling},
\begin{equation}
Z_{-} \rightarrow 0 \quad \Longrightarrow \quad B_* \rightarrow 0 \text {. }
\end{equation}
Interestingly, the same divergence that makes the unscreened $D_p^{-}$expensive also makes the discharge energetically favored, provided a finite-tension relative wall exists.

\section{Remarks}
Let us summarize the logic of the analysis. The starting point was the doubled RR sector of type \(0A\) and its interpretation, in the \(S^1_+\vee S^1_-\) picture of \cite{Baykara:2026bdv}, as data associated with two branches of the quantum wedge. In section \ref{section 4}, we used the tachyon equation of motion to isolate the branch-balance condition at \(T=0\): a symmetric point is naturally supported when the two RR sectors are balanced, while a branch-odd RR perturbation sources the tachyon fluctuation and measures an imbalance between the two DRP sectors. At the same time, for a \(Q\)-symmetric background the metric and dilaton are not sourced at first order by the branch-odd sector. This makes the relative RR sector a natural object to track as the system moves toward the tachyon-condensation endpoint. In section \ref{section 5}, we then asked what happens to a wrong-branch RR charge when the \(R_-\) branch collapses. The Gauss-law computation shows that an isolated unscreened \(D_p^-\) source has an electric self-energy scaling as \(1/Z_-\), so such an object becomes increasingly costly as \(Z_-\to0\), in direct analogy with the \(D0^-\) state becoming infinitely massive in the wedge picture. This motivates two logically distinct possibilities. The first is decoupling: the wrong-branch object simply leaves the low-energy Type IIA spectrum. The second is a stronger possibility: before the endpoint is reached, the object may discharge its relative type \(0A\) charge. To separate these possibilities we introduced two distinct notions. Screening refers to the disappearance of the long-range relative gauge field \(C_R=C^-_{p+1}-C^+_{p+1}\), for example through an effective higher-form Stückelberg mass; this explains why the relative charge \(Q_R\) is not visible in the Type IIA infrared theory. Discharge is even stronger: it requires a carrier of the relative charge \(\Delta(Q_+,Q_-)=(+1,-1)\), so that the charge vector can change from \((0,1)\) to \((1,0)\) while preserving the required IIA sector. The WZ/DBI construction above provides a natural effective representative of such a relative charge carrier, and the thin-wall estimate gives a parametric criterion for when the corresponding discharge channel is energetically favored. In this sense the overall goal is not to claim a microscopic derivation of the endpoint wall, but to give a coherent low-energy account of how the wrong-branch RR sector can be screened, and under what additional conditions it may discharge before the type IIA endpoint is reached.

We now spell out the regime in which this effective description should be interpreted. The WZ coupling fixes the relative charge carried by the fluxed object, while the DBI computation gives the expected normalization for the dissolved \(D_p\)-brane charge. In this sense the fluxed \(D(p+2)\)-brane provides a useful representative of the relative charge that must be carried by the interface. Strictly speaking, however, the object \(D(p+2)\) on \(\mathbb R^{p,1}\times S^2\), after reduction on the \(S^2\), has the same noncompact worldvolume dimension as a \(D_p\)-brane, whereas the Coleman wall in the \(D_p\)-brane worldvolume is a codimension-one interface with Euclidean geometry \(S^p\). We therefore regard the fluxed shell as capturing the charge carried by the wall, rather than as a complete microscopic construction of the Euclidean bounce geometry. Correspondingly, \(T_{\rm wall}\) should be treated as an effective wall tension, not as a quantity computed directly by the zero-radius DBI limit \(T_{\rm shell}^{\rm DBI}(0)=|n|T_p^{(0)}\). In the finite-tension regime, \(T_{\rm wall}\sim Z_-^0\), and when the pressure difference is dominated by the unscreened wrong-branch RR self-energy, \(\Delta\mathcal T\sim \mathcal E_-\sim 1/Z_-\), the thin-wall estimate gives \(B_*\sim Z_-^p\). More generally, allowing
\[
T_{\rm wall}\sim Z_-^{-a},
\qquad
\Delta\mathcal T\sim Z_-^{-\beta},
\]
one finds
\[
B_*\sim Z_-^{\beta p-a(p+1)}.
\]
Thus the discharge remains parametrically favored whenever the wall tension grows sufficiently slowly, \(a<\beta p/(p+1)\). Finally, the limit \(B_*\to0\) should be understood within the usual domain of validity of the thin-wall approximation: once the critical radius becomes comparable to the wall thickness or to the string scale, a more microscopic or thick-wall description is required. Similarly, the estimate \(\Delta\mathcal T\sim 1/Z_-\) applies in the regime where the wrong-branch charge is effectively unscreened on the relevant length scales; after the relative field \(C_R\) is fully screened or integrated out, the long-distance Coulomb estimate is replaced by short-distance core data. These qualifications do not affect the main point: the WZ/DBI analysis identifies the appropriate relative charge carrier, while the thin-wall computation gives a controlled parametric criterion for when such a discharge channel is energetically favored.
\section*{Acknowledgements}
ARK thanks Linus Wulff for comments on this draft, Thomas Van Riet for explaining and discussing Myers Effect and Brane-Flux Annihilation, Steven Hsia Weilong for discussions on higher form symmetries, Ratul Mahanta and Walid Hasan for discussions on Brane actions and Ashoke Sen for various conversations on dualities and tachyon condensation. ARK is supported by the Czech Science Foundation GA\v{C}R grant "One loop in ten and eleven dimensions"(GA26-22343S) and also acknowledges support from a BRAC University grant - BRACURSGI25014. ARK also thanks KU Leuven physics department and especially Thomas Van Riet for hosting at Leuven where part of this work was done. Finally, the title of the paper was inspired by a recent movie directed by Imtiaz Ali.

\appendix

\bibliographystyle{utphys}
\bibliography{biblio}

\end{document}